\pdfoutput=1
\documentclass[preprint,12pt]{elsarticle}

\usepackage{graphics}
\usepackage{graphicx}
\usepackage{epstopdf}
\usepackage{subfigure}
\usepackage{amssymb}
\usepackage{amsmath}

\journal{Complexity}

\biboptions{sort&compress}

\makeindex

\begin{document}
\begin{frontmatter}

\title{Dynamical variety of shapes in financial multifractality}

\author[address1,address2]{Stanis{\l}aw Dro\.zd\.z\corref{cor1}}
\cortext[cor1]{Corresponding author:}
\ead{stanislaw.drozdz@ifj.edu.pl}
\author[address1]{Rafa{\l} Kowalski}
\author[address1]{Pawe{\l} O\'swi\c ecimka}
\author[address1,address3]{Rafa{\l} Rak}
\author[address2]{Robert G\c ebarowski}

\address[address1]{Complex Systems Theory Department, Institute of Nuclear Physics, Polish Academy of Sciences, ul.~Radzikowskiego 152, 31-342 Krak\'ow, Poland}
\address[address2]{Faculty of Physics, Mathematics and Computer Science, Cracow University of Technology, ul.~Warszawska 24, 31-155 Krak\'ow, Poland}
\address[address3]{Faculty of Mathematics and Natural Sciences, University of Rzesz\'ow, ul.~Pigonia 1, 35-310 Rzesz\'ow, Poland}

\begin{abstract}
The concept of multifractality offers a powerful formal tool to filter out multitude of the most relevant characteristics of complex time series. The related studies thus far presented in the scientific literature typically limit themselves to evaluation of whether or not a time series is multifractal and width of the resulting singularity spectrum is considered a measure of the degree of complexity involved. However, the character of the complexity of time series generated by the natural processes usually appears much more intricate than such a bare statement can reflect. As an example, based on the long-term records of S\&P500 and NASDAQ - the two world leading stock market indices - the present study shows that they indeed develop the multifractal features, but these features evolve through a variety of shapes, most often strongly asymmetric, whose changes typically are correlated with the historically most significant events experienced by the world economy. Relating at the same time the index multifractal singularity spectra to those of the component stocks that form this index reflects the varying degree of correlations involved among the stocks. 
\end{abstract}

\end{frontmatter}

\section{Introduction}
\label{Intro}

Multifractality is a concept that is central to the science of complexity. The related multi-scale approach~\cite{halsey1986,mandelbrot1989,muzy1994} aims at bridging the wide range of time and length scales that are inherent in a number of complex natural phenomena and, as such, it pervades essentially all scientific disciplines~\cite{kwapien2012}. By now it finds applications in essentially all areas of the scientific activity, including physics~\cite{muzy2008,subramaniam2008}, biology~\cite{ivanov1999,makowiec2009,rosas2002}, chemistry~\cite{stanley1988,udovichenko2002}, geophysics~\cite{witt2013,telesca2005}, hydrology~\cite{koscielny2006}, atmospheric physics~\cite{kantelhardt2006}, quantitative linguistics~\cite{ausloos2012,drozdz2016}, behavioral sciences~\cite{ihlen2013}, cognitive structures~\cite{dixon2012}, music~\cite{jafari2007,oswiecimka2011}, songbird rhythms~\cite{roeske2018}, physiology~\cite{ghosh2014,nagy2017}, human behaviour~\cite{stephen2012,you2015,domanski2017}, social psychology~\cite{stephen2017} and even ecological sciences~\cite{stephen2013}, but especially frequently in economic and in financial contex~\cite{ausloos2002,calvet2002,turiel2005,oswiecimka2005,zhou2009,bogachev2009,su2009,drozdz2010,
oswiecimka2013,dutta2016,grech2016,bayraci2018} as stimulated by practical aspects and by needs to develop models of the financial dynamics based on multifractality~\cite{bacry2001,calvet2002,oswiecimka2006b,lux2008,kutner2008} such that they help in making predictions. Indeed, the multifractal analyses of the financial time series have provided so far most of the quantitative evidence for the factors that induce the genuine multifractality, such as the temporal long-range non-linear correlations and, only when such correlations are present, the fat tails in the distribution of fluctuations~\cite{zhou2012}. In order to unambiguously identify action of such factors and to suppress potential spurious multifractality the time series under study have to be, however, sufficiently long~\cite{drozdz2009}. In addition, the realistic time series, as generated by the natural phenomena, even if of multifractal character, are typically more involved in composition than the model mathematical uniform multifractals and they may contain several components of different multifractality characteristics. In such frequent cases the global hierarchical organization of the series gets distorted and the multifractal spectrum becomes asymmetric, either left- or right-sided, as recently demonstrated in ref.~\cite{drozdz2015}. Detecting such effects may provide even more valuable information about the mechanism that governs dynamics of a particular time series than just a bare statement that it is multifractal. Such effects of asymmetry are for instance, already found to constitute a very helpful formal tool in identifying a specific organization of complex networks~\cite{oswiecimka2018}. Furthermore, directions of the relevant distortions may vary in time parallel to changes of weight of the constituent components in a series. The most straightforward candidate to experience this kind of impact is the stock market index which, by construction, is already a sum, most often weighted, of prices of the constituent companies and those companies themselves may react differently for the same external news depending of the sector they belong to. It is primarily for this reason that below the world largest stock market indices are studied. Of course another, more specific, market oriented reason for this study is to broaden our historical perspective on evolution of the stock market multi-scale characteristics over periods comprising the global crashes or transitions due to the technological revolution in trading.

\section{Multifractal formalism \label{Method}}

At present there exist two distinct, commonly accepted and complementary computational methods that serve quantification of the multifractal characteristics of the time series. One of them - the Wavelet Transform Modulus Maxima (WTMM)~\cite{muzy1994} - makes use of the wavelet expansion of the time series under consideration and the other one - the Multifractal Detrended Fluctuation Analysis (MFDFA)~\cite{kantelhardt2002} - is based on inspecting the scaling properties of the varying order moments of fluctuations evaluated after an appropriate trend removal. While the former of those techniques allows a better visualisation of the underlying patterns in the time series the latter one often appears more accurate and more stable numerically and it will therefore be used here. Furthermore, at present there exists a consistent generalisation of MFDFA such that it even allows to properly identify and quantify the multifractal aspects of cross-correlations between two time series~\cite{podobnik2008,zhou2008,oswiecimka2014}. This novel method, termed Multifractal Cross-Correlation Analysis (MFCCA), consists of several steps that at the beginning are common to all the methods based on detrending.

One thus considers two time series $x_i$, $y_i$, where $i=1,2...T$. The signal profile is then calculated for each of them:
\begin{equation}
X(j) =\sum_{i=1}^j[x_{i}-\langle x\rangle] ,\quad
Y(j) =\sum_{i=1}^j[y_{i}-\langle y\rangle],
\end{equation}
where $\langle \rangle$ denotes averaging over the entire time series. Next, both these signal profiles are split up into $2M_s$ ($M_s=int(T/s)$) disjoint segments $\nu$ of length $s$ starting both from the beginning and the end of the profile and in each $\nu$, the assumed trend is estimated by fitting a polynomial of order $m$ ($P^{(m)}_{X,\nu}$ for $X$ and $P^{(m)}_{Y,\nu}$ for $Y$). In typical cases an optimal choice corresponds to $m=2$~\cite{oswiecimka2006}. This trend is subtracted from the series and the detrended cross-covariance within each segment is calculated:
\begin{multline}
F_{xy}^{2}(\nu,s)=\frac{1}{s}\Sigma_{k=1}^{s}\lbrace
(X((\nu-1)s+k)-P^{(m)}_{X,\nu}(k)) \times \\ \times
(Y((\nu-1)s+k)-P^{(m)}_{Y,\nu}(k))\rbrace .
\label{Fxy2}
\end{multline}
Since $F_{xy}^{2}(\nu,s)$ can assume both positive and negative values the $q$th order covariance function is defined by the following equation:
\begin{equation}
F_{xy}^{q}(s)=\frac{1}{2M_s}\Sigma_{\nu=1}^{2M_s} {\rm
sign}(F_{xy}^{2}(\nu,s))|F_{xy}^{2}(\nu,s)|^{q/2} ,
\label{Fq}
\end{equation}
where ${\rm sign}(F_{xy}^{2}(\nu,s))$ denotes the sign of $F_{xy}^{2}(\nu,s)$.
The parameter $q$ in Eq.~(\ref{Fq}) can take any real number except zero. However, for $q=0$, the logarithmic version of this Equation can be employed~\cite{kantelhardt2002}:
\begin{equation}
F_{xy}^{0}(s)=\frac{1}{2M_s}\Sigma_{\nu=1}^{2M_s} {\rm sign}(F_{xy}^{2}(\nu,s))
ln|F_{xy}^{2}(\nu,s)| .
\label{Fqln}
\end{equation}

Fractal cross-dependences between the time series $x_i$ and $y_i$ then manifest themselves in the scaling relations:
\begin{equation}
F_{xy}^{q}(s)^{1/q}=F_{xy}(q,s) \sim s^{\lambda _q}
\label{Fxy}
\end{equation}
(or $\exp(F_{xy}^{0}(s)) = F_{x,y}(0,s) \sim s^{\lambda_0} $ for $q=0$), where
$\lambda_q$ is the corresponding scaling exponent whose range of dependence on $q$ quantifies the degree of the complexity involved. Scaling with the $q$-dependent exponents reflects a richer, multifractal character of correlations in the time series as compared to monofractal case when $\lambda_q$ is $q$-independent.

The conventional MFDFA procedure of calculating the singularity spectra for single time series can be considered a special case of the above MFCCA procedure and corresponds to taking $x_i$, $y_i$ as identical. The Eq.~(\ref{Fq}) then reduces to:
\begin{equation}
F(q,s)=\Big[\frac{1}{2M_s}\sum^{2M_s}_{\nu=1}{[F^2(\nu,s)]^{\frac{q}{2}}}\Big]^{\frac{1}{q}}
\label{F}
\end{equation}
and to a corresponding counterpart of Eq.~(\ref{Fqln}) for $q=0$. The signatures of multifractality (monofractality) are then reflected, analogously to Eq.~(\ref{Fxy}), by
\begin{equation}
F(q,s) \sim s^{h(q)},
\end{equation}
where $h(q)$ denotes the generalized Hurst exponent. The singularity spectrum (also referred to as multifractal spectrum) $f(\alpha)$ is then calculated from the following relations:
\begin{equation}
\alpha=h(q)+qh'(q), \quad f(\alpha)=q[\alpha-h(q)]+1 ,
\end{equation}
where $\alpha$ denotes the H\"older exponent characterizing the singularity strength and $f(\alpha)$ reflects the fractal dimension of support of the set of data points whose H\"older exponent equals $\alpha$. In the case of multifractals the shape of the singularity spectrum typically resembles an inverted parabola and the degree of their complexity is straightforwardly quantified by the width of $f(\alpha)$:
\begin{equation}
\Delta \alpha = \alpha _{max}  - \alpha _{min} ,
\end{equation}
where $\alpha _{min}$ and $\alpha _{max}$ correspond to the opposite ends of the $\alpha$ values as projected out by different $q$-moments (Eq.~(\ref{F})). For monofractal signals the spectrum converges to a single point, though in practice this often turns out to be a subtle matter~\cite{drozdz2009}. Another important feature of the multifractal spectrum is its asymmetry (skewness), which can be quantified by the asymmetry coefficient \cite{drozdz2015}:
\begin{equation}
A_{\alpha}=\frac{\Delta \alpha _L - \Delta \alpha _R}{\Delta \alpha _ L + \Delta \alpha _R},
\end{equation}
where $\Delta \alpha _L = \alpha _0 - \alpha _{min}$ and $\Delta \alpha _R = \alpha _{max} - \alpha _0$ and for $\alpha _0$ the spectrum $f(\alpha)$ assumes maximum. The positive value of $A _\alpha$ reflects the left-sided asymmetry of $f(\alpha)$, i.e. its left arm is stretched with respect to the right one, and thus more developed multifractality on the level of large fluctuations in the time series. Negative $A _\alpha$, on the other hand, reflects the right-sided asymmetry of the spectrum and indicates temporal organization of the smaller fluctuations as the main source of multifractality.

A family of the fluctuation functions as defined by Eq.~(\ref{Fq}) can also be used to define a $q$-dependent detrended cross-correlation ($q$DCCA)~\cite{kwapien2015} coefficient
\begin{equation}
\rho_q(s) = {F_{xy}^q(s) \over \sqrt{ F_{xx}^q(s) F_{yy}^q(s) }},
\label{rho.q}
\end{equation}
which allows to quantify the degree of cross-correlations between two time series $x_i$, $y_i$ after detrending and at varying time scales $s$. Furthermore, by varying the parameter $q$ one is able to identify the range of detrended fluctuation amplitudes that are correlated most in the two signals under study~\cite{kwapien2015}. This filtering ability of $\rho_q(s)$ constitutes an important advantage as cross-correlations among time series typically are not uniformly distributed over their fluctuations of different magnitude~\cite{kwapien2017}.

\section{Data specification}
\label{Data}

In the present study two sets of data are used:
\begin{itemize}
\item Daily prices of the S\&P500 and NASDAQ indices covering the period January 03, 1950 - December 29, 2016 (16496 data points). The values of NASDAQ before 1971 (official launching date of the index is February 05, 1971) were reconstructed from the historical data \cite{data}.
\item Daily prices of 9 stocks listed on the NYSE over the period from January 1, 1962 to July 07, 2017 (13812 points). The analysed companies are GE - General Electric, AA - Alcoa, IBM - International Business Machines, KO - Coca Cola, BA -  Boeing, CAT - Caterpillar, DIS - Walt Disney, HPQ - Hewlett Packard, DD - DuPont. These in fact are the only stocks that participate in the Dow Jones Industrial Average (DJIA) over such a long period of time, and thus also in S\&P500. They, however, represent a large spectrum of the economy sectors and may thus be considered as a reasonable representation for the larger American indices.
\end{itemize}
For each time series the logarithmic returns are calculated according to the equation:
\begin{equation}
r(t+\Delta t)=\ln p(t+\Delta t) -\ln p(t),
\end{equation}
where $p(t)$ denotes the stock price or index value and $\Delta t$ stands for time interval ($\Delta t$ = 1 day). All time series are normalized to have unit variance and zero mean.

\section{Results}
\label{results}

\subsection{S{\normalfont{\&}}P500 and NASDAQ}
\label{twoindices}

The MFDFA multifractal spectra $f(\alpha)$ for the S\&P500 and NASDAQ indices are shown in Fig.~1. For both these indices the fluctuation functions $F(q,s)$ reveal a convincing power law behaviour over almost two decades, which is shown in the corresponding lower-right insets thus $f(\alpha)$ is determined unambiguously. The parameters $q$ are taken within the interval $-4 \le q \le 4$, which is common in financial applications because it allows to safely avoid the danger of divergent moments when the fluctuation functions $F(q,s)$ are computed. Cumulative distributions of the return fluctuations for the two indices considered here are shown in the corresponding upper-left panels of Fig.~1 and can be seen not to develop thicker tails than the inverse cubic power-law~\cite{gopi1998,drozdz2007} and there is thus no danger of the divergent moments. The width of the resulting spectra $\Delta \alpha \approx 0.4$ for S\&P500 and $0.32$ for  NASDAQ, correspondingly. The significance of this result is also tested against the two null hypotheses of $f(\alpha)$ calculated from (i) the series obtained from the original ones by a random shuffling, thus destroying all the temporal correlations (green triangles) and (ii) Fourier-phase randomised counterparts of the original series which destroys the nonlinear correlations (blue squares). Clearly, the $f(\alpha)$ spectra in these two tests get shrank to a form characteristic to monofractals. An additional form of surrogates tested here are time series with the Gaussianized pdf's. In the latter case the original pdf is replaced by a Gaussian distribution while the amplitude ranks of fluctuations remain preserved. The resulting multifractal spectra appear only slightly narrower than the original ones and therefore they are not shown in Fig.~1. All these tests thus provide a convincing evidence for quite a rich multifractality of the original time series and, moreover, corroborate the fact that this multifractality is, as expected~\cite{drozdz2009}, due to the nonlinear temporal correlations. The obtained multifractal spectra are at the same time visibly left-sided asymmetric~\cite{drozdz2015}. The asymmetry coefficient $A_\alpha \approx 0.3$ for S\&P500 and $0.31$ for NASDAQ. The left side of $f(\alpha)$ is determined by the positive $q$-values which filter out larger events and the opposite applies to the right side of this spectrum. In the present context this thus means that it is the dynamics of the large returns which develops more pronounced multifractal organization than that of the small returns.

\begin{figure}
\centering
\includegraphics[scale=0.49,keepaspectratio=true]{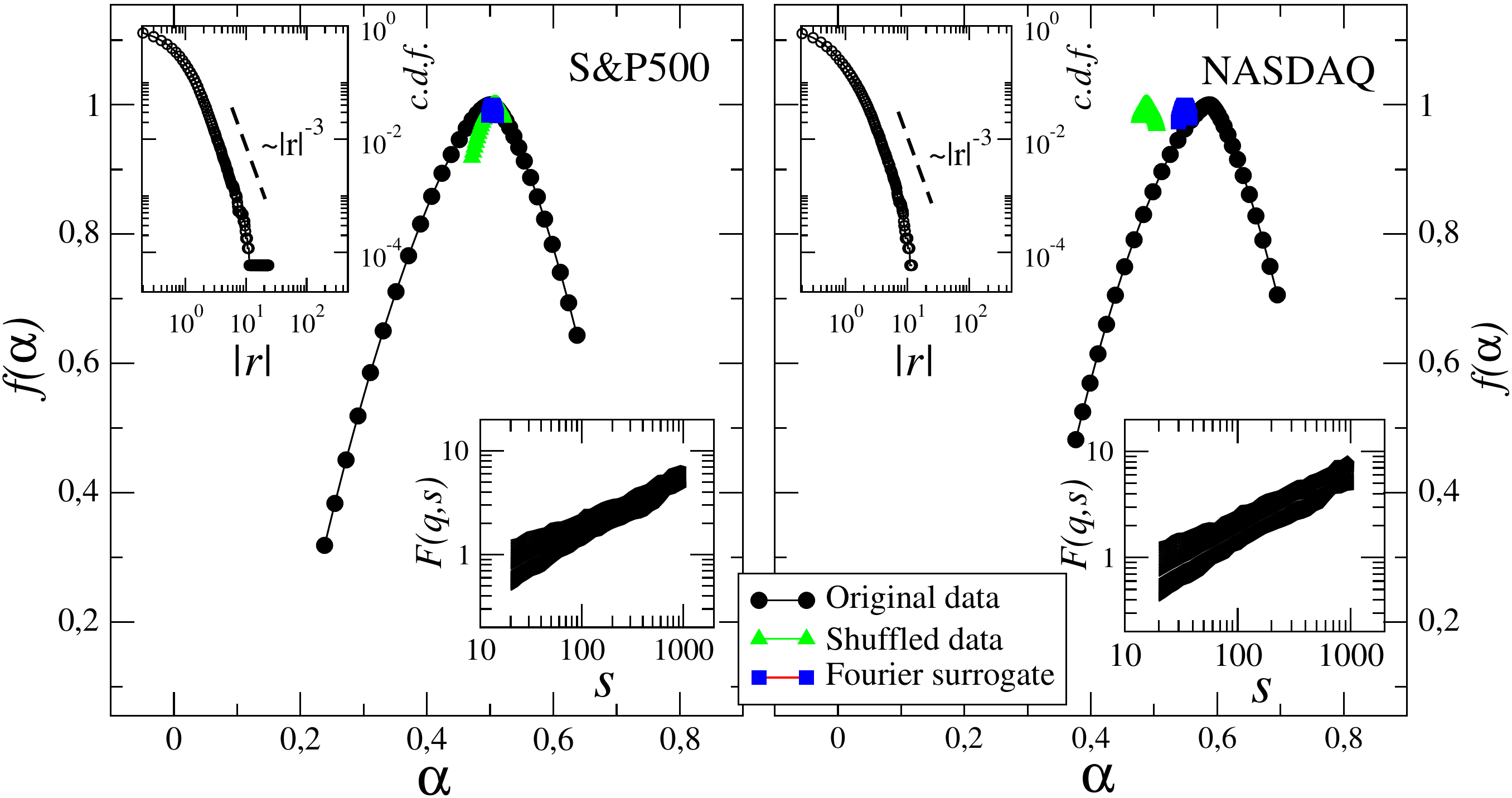}
\caption{Main panels: Multifractal spectra calculated for the S\&P500 and NASDAQ returns (black dots) covering the period January 03, 1950 - December 29, 2016. Average spectra obtained for the Fourier phase-randomized  surrogates and for the randomly shuffled time series are denoted by blue squares and green triangles, respectively. Upper-left insets display cumulative distributions of return fluctuations and lower-right insets display the fluctuation functions calculated for the original S\&P500 and NASDAQ series.}
\label{Fig1}
\end{figure}

Fig.~1 shows the result of calculations over the entire time span where the time series are taken. It appears that probing this period with a shorter window rolling in time reveals a non-trivial and a very interesting time-dependence of the corresponding multifractal spectra. Here the window size is taken over 5000 data points (equivalent to about 20 years) which in presence of temporal correlations is sufficiently long to guaranties stability of the result~\cite{drozdz2009} (absence of such correlations demands significantly longer series~\cite{drozdz2009}) and the window is moved with the step of 20 points (approximately one calendar month). The results of such a procedure are highlighted in Fig.~2 for the S\&P500 and in Fig.~3 for the NASDAQ. Panels (a) in these Figures show sequences of the singularity spectra $f(\alpha)$, calculated within such windows consecutively, and the calendar date assigned to each $f(\alpha)$ corresponds to the end point within a window. Thus, for the time series which begin, as here, in January 1950 the first date appearing in Figs.~2 and 3 corresponds to January 1969. In order to better visualize evolution of $\Delta \alpha$ and of $A_{\alpha}$ the panels (b) in these Figures show projections of $f(\alpha)$ onto the time $(t)$ - $\alpha$ plane. The three historically most recognized events that influenced the world financial markets are indicated by the vertical dashed lines. These are the Black Monday of October 19, 1987, burst of the Dot-com bubble in March 10, 2000, and bankruptcy of the Lehman Brothers in September 15, 2008.  Clearly, evolution of $f(\alpha)$ in such a 20-years time-window reveals sizeable changes in the width of $f(\alpha)$ and in its asymmetry, both going somewhat differently in the two indices, however. For S\&P500 until about 1985 the spectrum is comparatively  broad and then starts quick narrowing but this narrowing primarily results from shrinkage of the right arm in $f(\alpha)$. For the time-window ending in around 1993 this arm almost disappears and starts recovering only in recent years. Interestingly, the left side of $f(\alpha)$ got broadened even a few years earlier. The NASDAQ spectrum $f(\alpha)$ also experiences sizeable changes in time but differently than the one for S\&P500. On average this spectrum is broader and strongly asymmetric for time windows ending between about Black Monday and the burst of Dot-com bubble in 2000 but here this asymmetry results from a sudden stretching of the left side of $f(\alpha)$ while the right side does not experience much changes.

\begin{figure}
\centering
\includegraphics[scale=0.11,keepaspectratio=true]{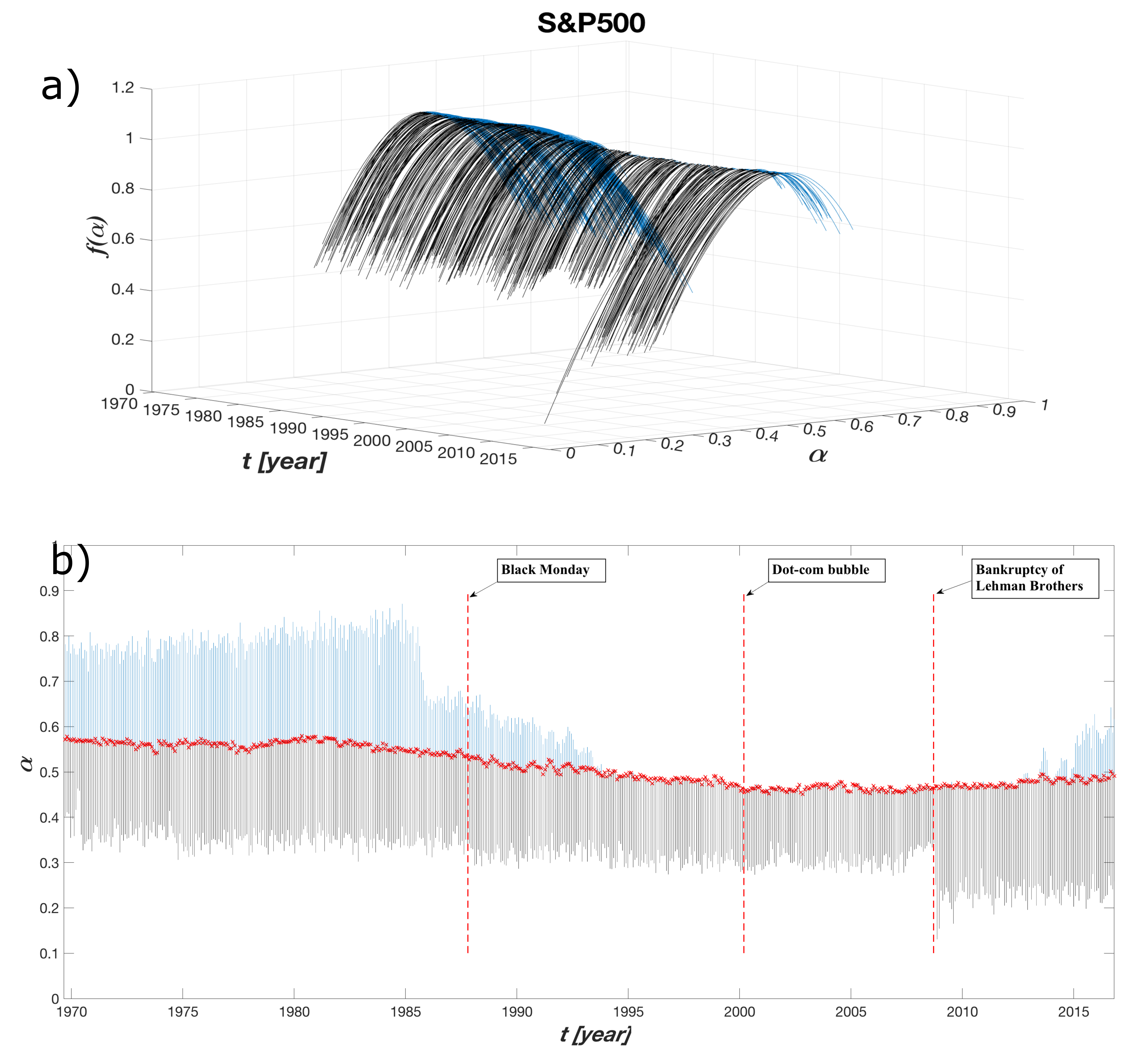}
\caption{Panel (a): For S\&P500 from January 03, 1950 to December 29, 2016 the sequence of singularity spectra $f(\alpha)$ calculated within a rolling 20-years window. The calendar date assigned to each $f(\alpha)$ corresponds to the end point within a window. This window is moved with the step of 20 points which corresponds approximately to one calendar month. Black sold line corresponds to the left and blue line to the right side of $f(\alpha)$. Panel (b): Projections of $f(\alpha)$ of Panel (a) onto the time $(t)$ - $\alpha$ plane. Red line illustrates displacement of the maxima of $f(\alpha)$ in the consecutive windows. Vertical dashed lines indicate the Black Monday of October 19, 1987, burst of the Dot-com bubble in March 10, 2000, and bankruptcy of Lehman Brothers in September 15, 2008.}
\label{Fig2}
\end{figure}

\begin{figure}
\centering
\includegraphics[scale=0.11,keepaspectratio=true]{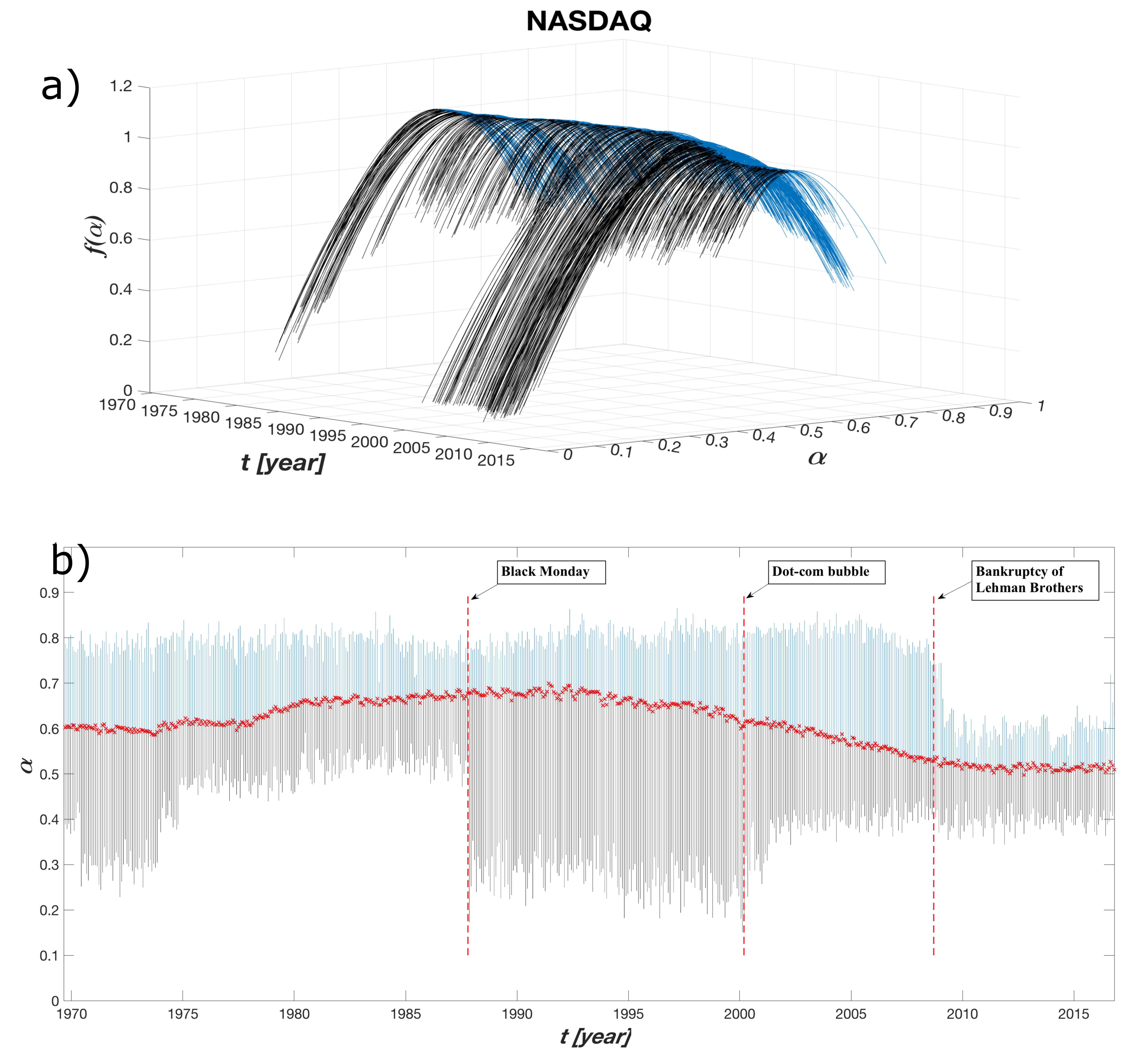}
\caption{As in Fig.~2 but analysis is carried out for the NASDAQ data.}
\label{Fig3}
\end{figure}

In Figs~2 and 3 one also sees changes in location of the maxima of $f(\alpha)$ which are related to a degree of persistence in times series. A parameter that directly quantifies this property is the Hurst exponent $H=h(2)$. Fig.~4 shows the Hurst exponents $H$, the widths $\Delta \alpha$, the asymmetry coefficients $A_{\alpha}$ and the widths ${\Delta \alpha}_{L(R)}$ for the time-sequence of the multifractal spectra already presented in Fig.~2 (S\&P500), whereas Fig.~5 shows these characteristics corresponding to Fig.~3 (NASDAQ). The two dates seen to be related to almost discontinuous changes in some of these quantities are the Black Monday of October 19, 1987 and the Bankruptcy of Lehman Brothers in September 15, 2008 and these two dates are indicated by the vertical dashed lines. While Black Monday affected the NASDAQ much more spectacularly than the S\&P500, though the latter started assuming similar trends already some 2 years earlier, the effect of Bankruptcy of Lehman Brothers was just the opposite. This time it is the S\&P500 which reveals a sudden increase of $\Delta \alpha$ by a factor of about 2 but, remarkably, this increase is entirely due to stretching of the left arm of $f(\alpha)$. A partial identification of the origin of these S\&P500 versus NASDAQ differences comes from Fig.~6, which displays fluctuations of the daily returns of these two indices and, as the most informative, the time dependence of the local (in the rolling window of $s=500$ trading days) detrended variance. In around the Black Monday this variance is much larger for NASDAQ than for S\&P500 and this goes in parallel with a sharp stretching of the left arm in $f(\alpha)$ for NASDAQ. On the other hand, in around the Bankruptcy of Lehman Brothers, even though the NASDAQ detrended variance still is somewhat larger than the one of S\&P500, it is much smaller than around the period of the Dot-com burst.
In the S\&P500 case the corresponding development is just reversed and larger variance accompanies the bankruptcy of Lehman Brothers. Thus, in this latter period the detrended variance of NASDAQ decreases while the one of S\&P500 increases and it is in this period when the left arm of $f(\alpha)$ for S\&P500 experiences a sudden stretching.
Worth noticing is also the fact that the Hurst exponents $H$ of these two indices on average decrease when going from past to present and in recent years assume values even lower than $0.5$, which indicates anti-persistence \cite{grech2004}. Especially monotonic in this respect is the S\&P500 - one of the most significant global indicators of the world economy - whose Hurst exponent on average systematically decreases over the whole time span considered and in the last couple of years it even steadily dropped down below 0.4. There are presumptions~\cite{grech2004,czarnecki2008} that such values of $H$ indicate proximity to a crash zone. In view of this result the log-periodic scenario~\cite{drozdz2003} indicating danger of a much larger world economic decline in around 2025 than anything the World has experienced so far needs to be taken into consideration more and more seriously. 

\begin{figure}
\centering
\includegraphics[scale=0.49,keepaspectratio=true]{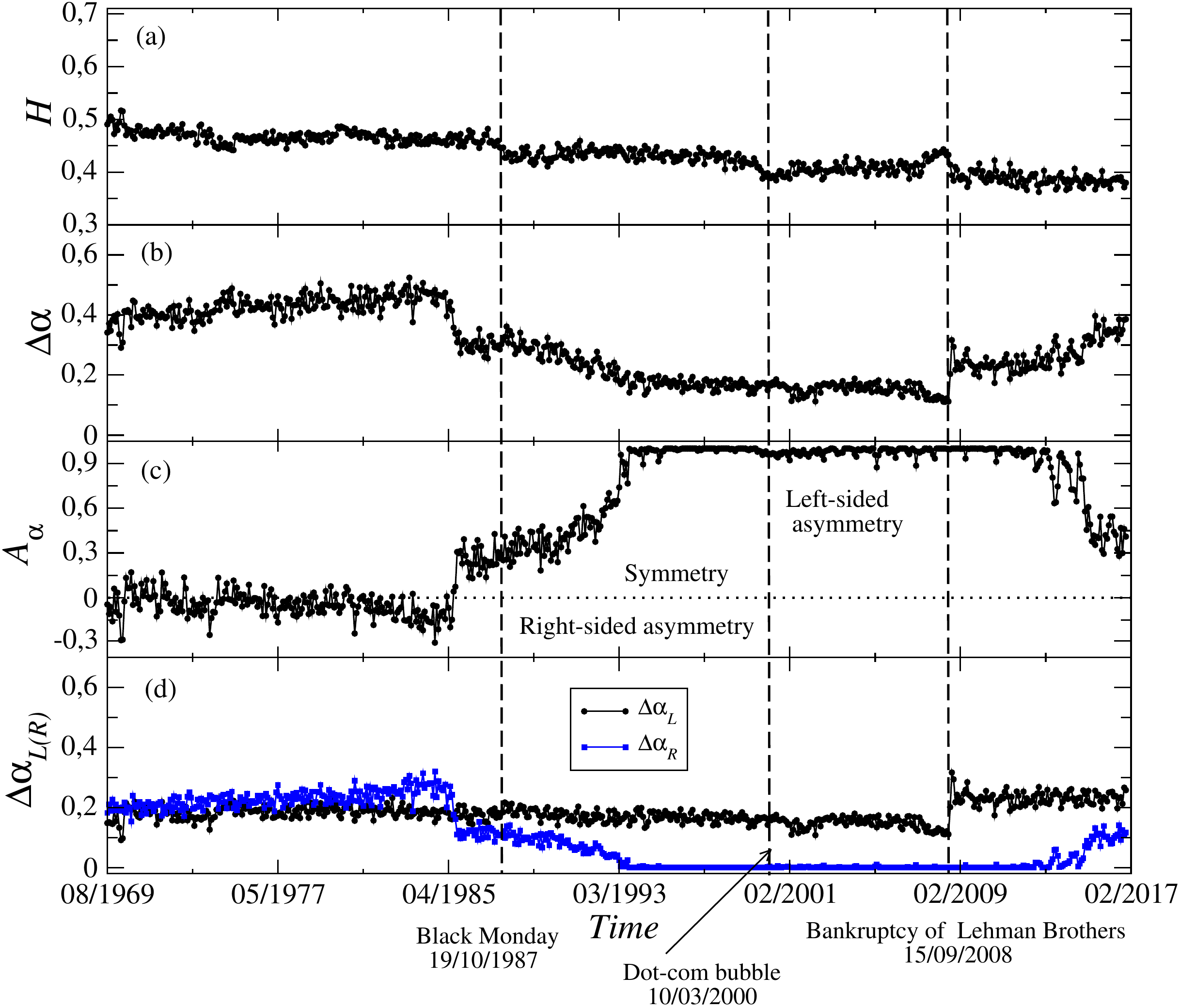}
\caption{For S\&P500: the Hurst exponents $H$, the widths $\Delta \alpha$, the asymmetry coefficients $A_{\alpha}$ and the widths ${\Delta \alpha}_{L(R)}$ for the time-sequence of multifractal spectra in 20-years windows of Fig.~2.}
\label{Fig4}
\end{figure}

\begin{figure}
\centering
\includegraphics[scale=0.49,keepaspectratio=true]{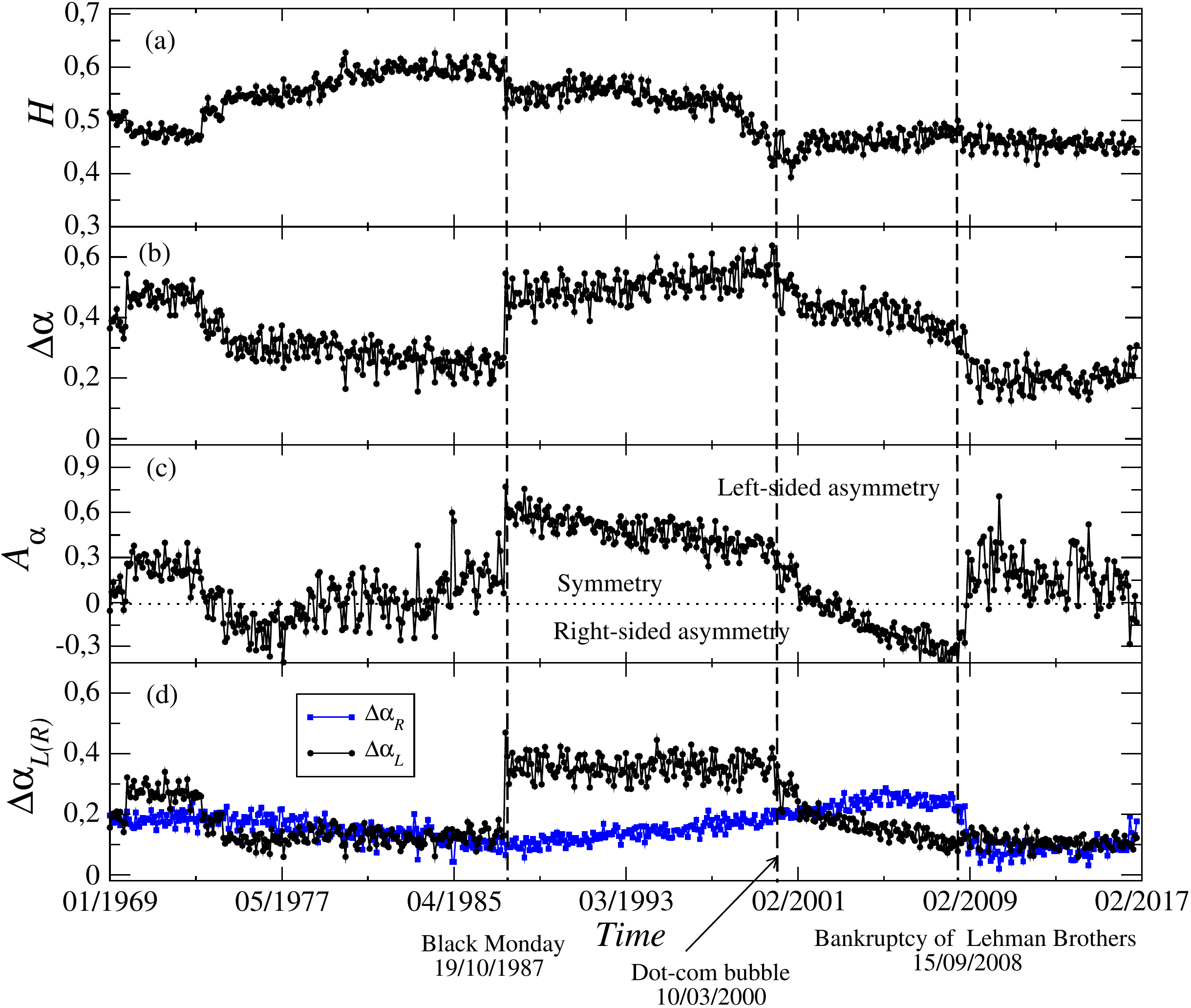}
\caption{For NASDAQ: the Hurst exponents $H$, the widths $\Delta \alpha$, the asymmetry coefficients $A_{\alpha}$ and the widths ${\Delta \alpha}_{L(R)}$ for the time-sequence of multifractal spectra in 20-years windows of Fig.~3.}
\label{Fig5}
\end{figure}

\begin{figure}
\centering
\includegraphics[scale=0.49,keepaspectratio=true]{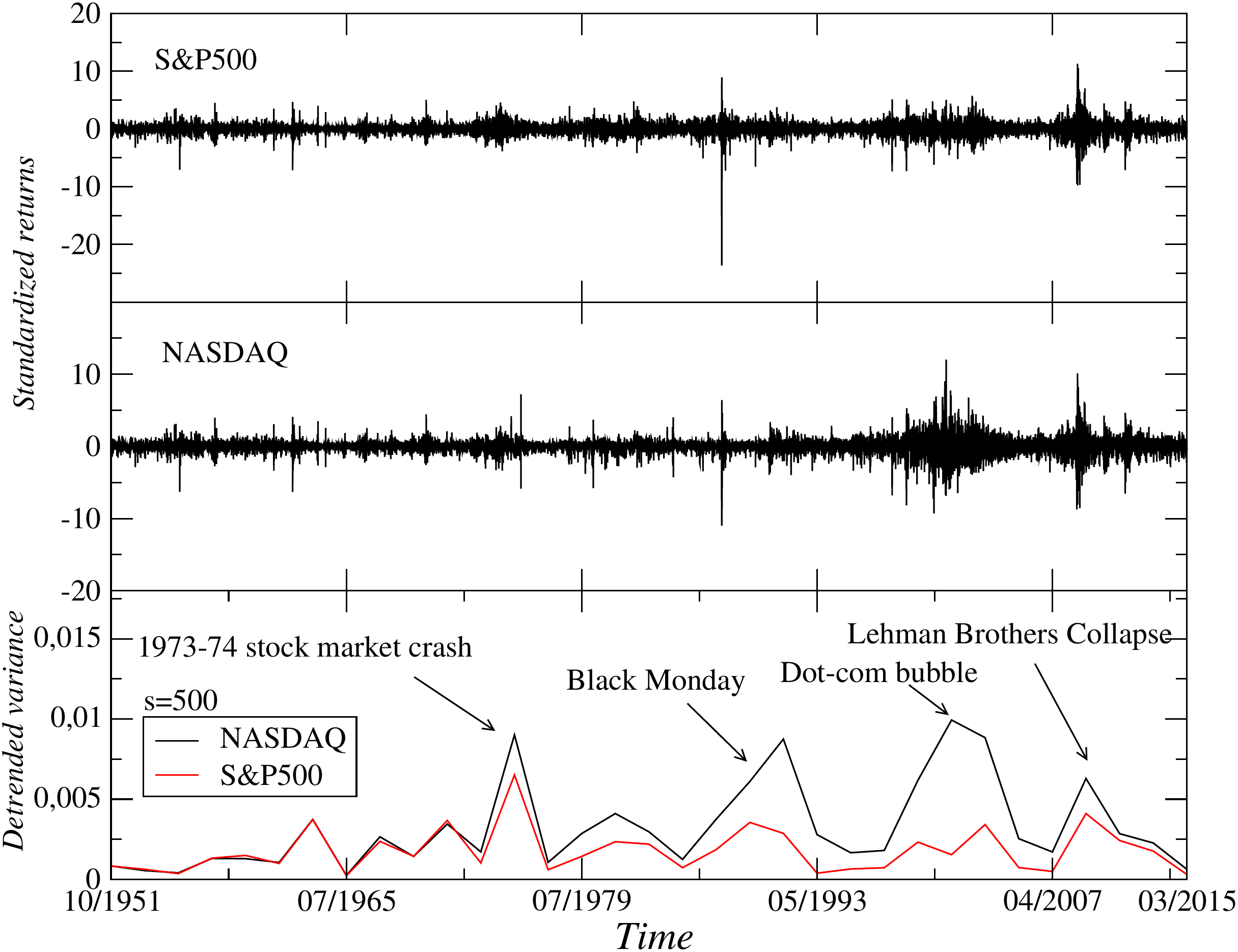}
\caption{Two upper panels: Daily returns for S\&P500 and for NASDAQ over the period January 03, 1950 - December 29, 2016. The bottom panel: the corresponding detrended variance for S\&P500 (red line) and for NASDAQ (black line).}
\label{Fig6}
\end{figure}

The window probed multifractal spectra of Figs.~2 and 3 for the S\&P500 and for the NASDAQ resemble each other more in the first half of the entire considered interval, until about mid 1980s, than in the following second half. This similarity or dissimilarity appears to occur even on the deeper level of their multifractal synchrony as reflected by the appropriate cross-correlations measures expressed by Eq.~\ref{Fxy}. The two approximately 20-years long time-periods taken from inside of these halves are selected as September 25, 1957 -- August 26, 1977 and May 19, 1989 -- March 20, 2008, the cross-correlations fluctuation functions between the S\&P500 and NASDAQ calculated according to Eq.~\ref{Fxy} and the result is shown in Fig.~7. It is very interesting to see that in the first of these periods the fluctuation functions display a clear tendency to scaling, which indicates cross-correlations between the two indices even on the level of their multifractal organization. This holds down to the level of their small fluctuations as measured by the negative $q$-values. In the second of these time-intervals, while for the positive $q$-values one may still see some remnants of scaling, for the negative $q$-values there is none, thus the indices are systematically loosing their multifractal synchrony and on the level of the small fluctuations this synchrony is lost completely.

\begin{figure}
\centering
\includegraphics[scale=0.49,keepaspectratio=true]{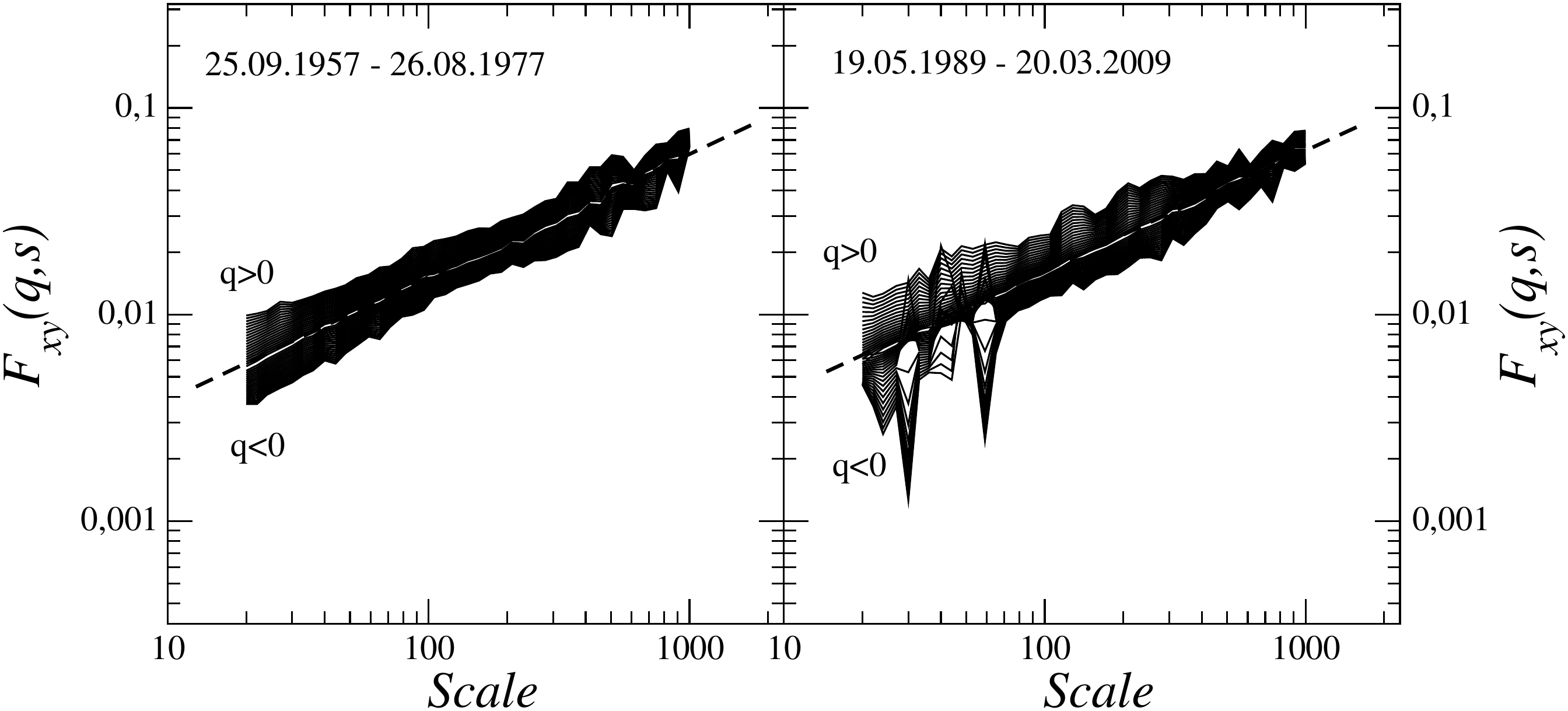}
\caption{Cross-correlations fluctuation functions between the S\&P500 and NASDAQ calculated according to Eq.~\ref{Fxy} for $-4 \leq q \leq 4$ in two periods: September 25, 1957 -- August 26, 1977 and May 19, 1989 -- March 20, 2009.}
\label{Fig7}
\end{figure}

\subsection{Index versus companies}
\label{IvsC}

It is natural to expect that significant changes in time of the multifractal features of the two indices seen in the previous subsection reflect different market phases and such phases vary in a degree of coupling among the component shares~\cite{drozdz2000}. These are the individual stocks which are traded and only a superposition of their multifractal characteristics, not necessarily identical, determines $f(\alpha)$ of an index. It is clear that in an uncorrelated sum of many multifracal time series the multifractality gradually disappears when the number of component series increases and, in addition, this limiting case is typically approached asymmetrically~\cite{drozdz2015}. One may thus anticipate that stronger coupling among the companies that form a basket of an index favours multifractality of that index as well. In the present context, in order to study such effects in more detail, by summing up prices of the 9 companies listed in Section~\ref{Data} a proxy of the DJIA is formed. It however, amazingly accurately follows changes in the full DJIA and even all the significant moves in the S\&P500, as it can be seen from Fig.~8. This is likely to reflect the fact that the 9 companies are dispersed over different market sectors and in total they well represent the global DJIA market.

\begin{figure}
\centering
\includegraphics[scale=0.49,keepaspectratio=true]{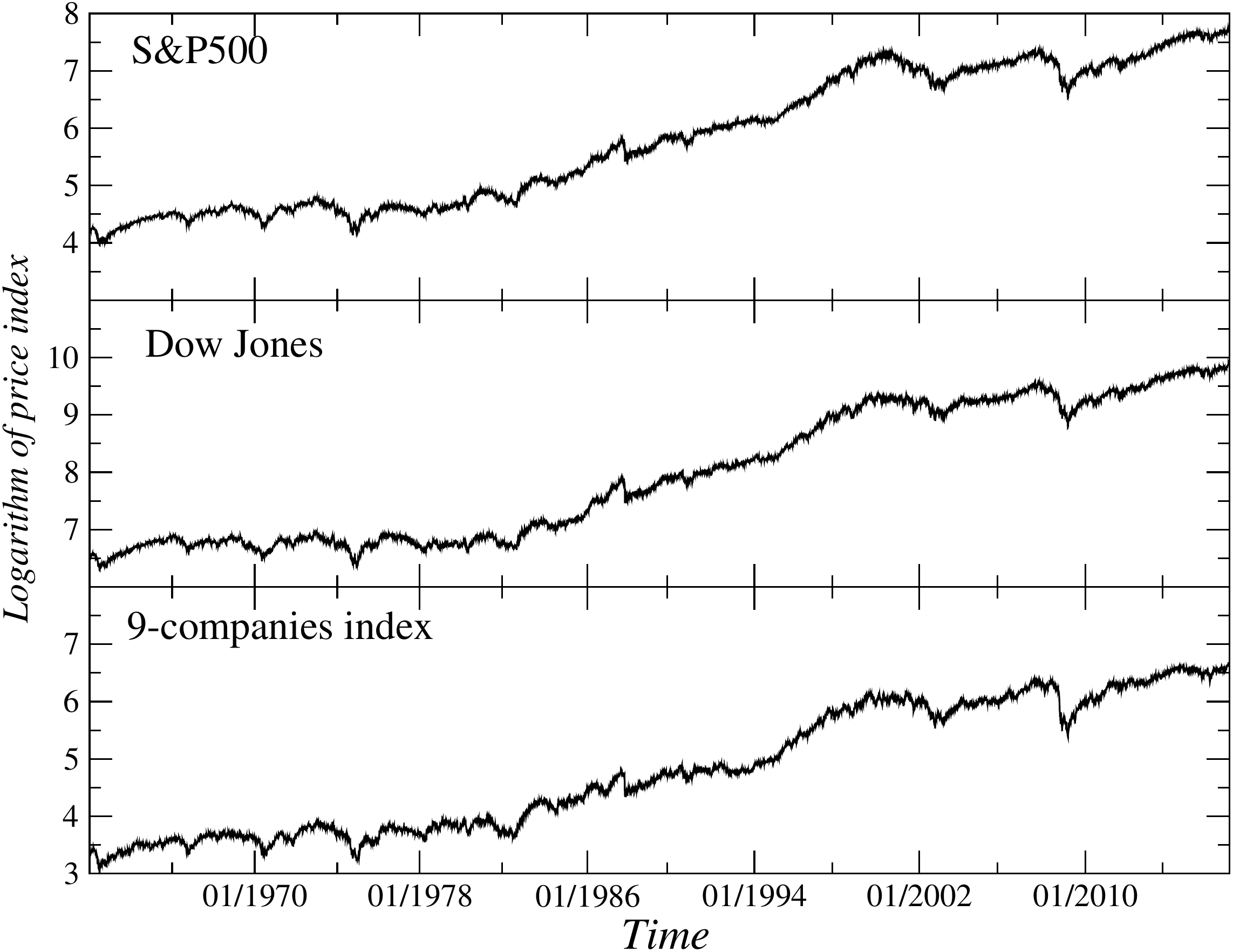}
\caption{Daily prices of S\&P500, of Dow Jones and of the sum of 9 DJIA stocks listed on the NYSE over the period from January 1, 1962 to July 07, 2017 (13812 points). The companies included are GE - General  Electric, AA - Alcoa, IBM - International Business Machines, KO - Coca Cola, BA -  Boeing, CAT - Caterpillar, DIS - Walt Disney, HPQ - Hewlett Packard, DD - DuPont.}
\label{Fig8}
\end{figure}

The results of calculations relating to the multifractal spectra $f_i(\alpha)$, projected onto the time $(t)$ - $\alpha$ plane, of these $N=9$ companies labelled by $i$ (thus here $i=1,...,9$), for illustrative clarity represented by one average $\tilde f(\alpha)= N^{-1} \sum_{i=1}^{i=N} f_i(\alpha)$ and of the index constructed from these 9 companies, in the same rolling window as before, are displayed in panels (a) and (b) of Fig.~9, correspondingly. Several interesting observations based on these results can be made. One main finding is that the width $\Delta \alpha$ of $\tilde f$ is never smaller than that of the global 9-companies index, which is understandable because equality is expected in the case of perfect correlation among prices of all the participating companies. Some decorrelation, which is always the case in real markets, should result in narrowing $f(\alpha)$ of the global, here 9-companies, index. A significantly larger difference between the widths of multifractal spectra in the two cases considered is observed for the time-period between the Black Monday and the Bankruptcy of Lehman Brothers and the transition is nearly sharp. This difference originates, however, from a sudden stretching of the left side in $\tilde f(\alpha)$ within that period, which indicates that multifractality of the price changes of individual companies is much more pronounced on the level of larger fluctuations than on the level of small ones. When prices of these companies are summed up to form a global 9-companies index this huge left side stretching is significantly reduced, which indicates that the large fluctuations of individual stocks are not fully correlated among themselves. Still, within this most volatile period (Fig.~\ref{Fig6}) in the market even the global index preserves the left-sided asymmetry in $f(\alpha)$ indicating dominance of non-linear correlations on the level of large fluctuations.

\begin{figure}
\centering
\includegraphics[scale=0.11,keepaspectratio=true]{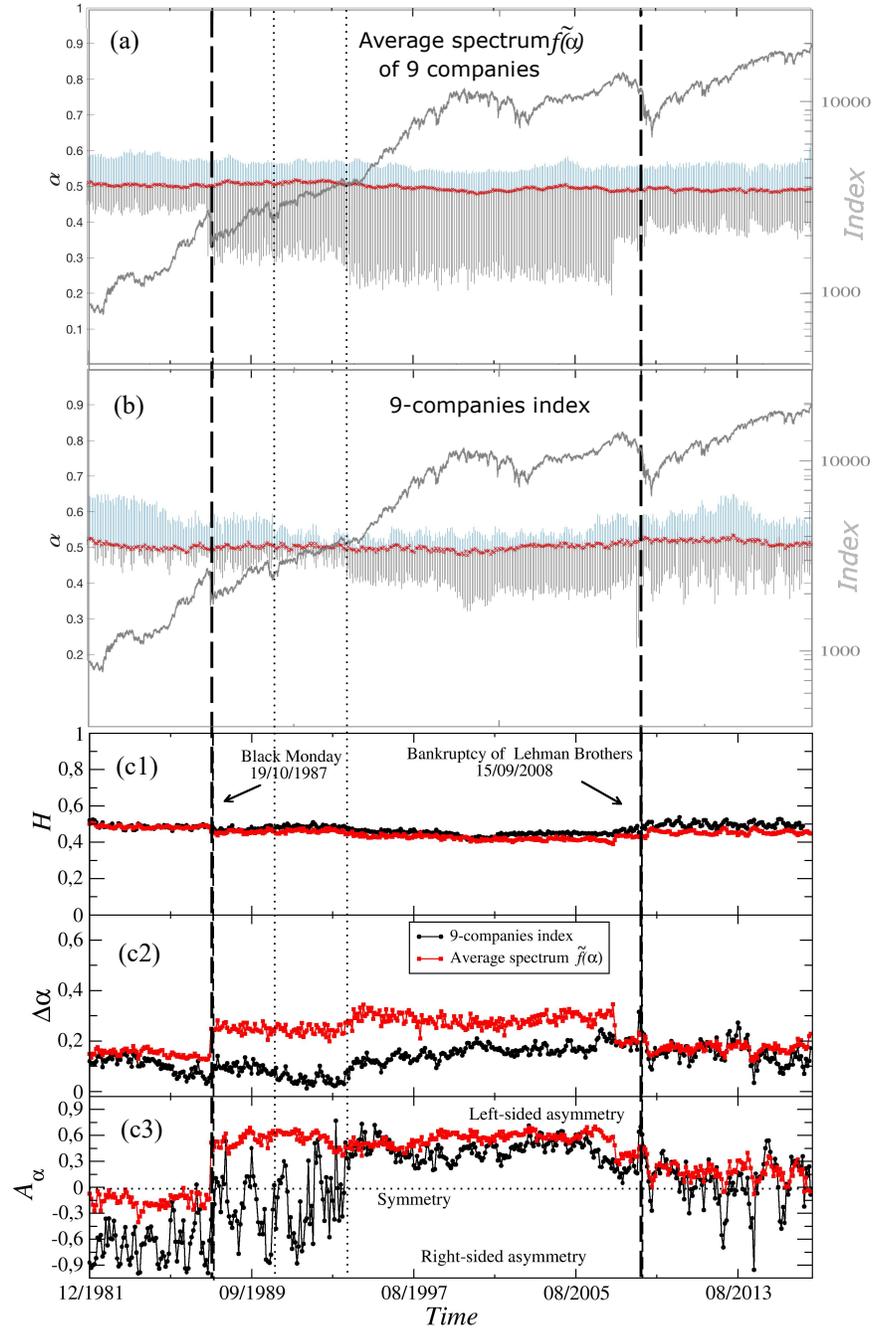}
\caption{Projections onto the time $(t)$ - $\alpha$ plane of the sequence of singularity spectra $f(\alpha)$ calculated within a rolling 20-years window for both the average spectrum $\tilde f(\alpha)$ (a) as well as of the artificial index of 9-companies of Fig.~\ref{Fig8} (b). The calendar date assigned to each $f(\alpha)$ corresponds to the ending point within a window. This window is moved with the step of 20 points which corresponds to approximately one calendar month. Red line illustrates displacement of the maxima of $f(\alpha)$ in the consecutive windows. Vertical dashed lines indicate the Black Monday of October 19, 1987 and bankruptcy of the Lehman Brothers in September 15, 2008 while the dotted ones October 1990 and April 1994. The bottom three panels display the corresponding Hurst exponents $H$ (c1), widths $\Delta \alpha$ (c2) and the asymmetry coefficients $A_{\alpha}$ (c3).}
\label{Fig9}
\end{figure}

An especially interesting related case occurs in the period between October 1990 and April 1994 indicated in Fig.~9 by the two vertical dotted lines. In this period the multifractal spectra of the individual companies on average develop broad multifractal spectra while $f(\alpha)$ of the corresponding global 9-companies index is so narrow that it can be considered as monofractal. One possible reason for such a result is a substantial suppression of cross-correlations among price changes of the component stocks~\cite{drozdz2015}.

Such a possibility is verified using the correlation matrix
\begin{equation}
{\bf C} = (1/ T) \ {\bf M} {\bf M}^{\bf T},
\label{C}
\end{equation}
where $\bf M$ denotes a $N \times T$ rectangular matrix formed from $N$ time series $x_i(t)$ of length $T$. Entries of the matrix ${\bf C}$ thus correspond to the conventional Pearson correlation coefficients. By diagonalizing $\bf C$ $({\bf C} {\bf v}^k = \lambda_k {\bf v}^k)$ one obtains the eigenvalues $\lambda_k$ $(k=1,...,N)$ and the corresponding eigenvectors ${\bf v}^k$. In the limiting case of entirely random signals the density of eigenvalues $\rho_C(\lambda)$ is known analytically~\cite{marchenko1967,edelman1988} as
\begin{equation}
\rho_C(\lambda) = {Q \over {2 \pi\sigma^2}} {\sqrt{ (\lambda_{max} - \lambda) (\lambda -\lambda_{min})} \over {\lambda}},
\end{equation}
where the lower $\lambda_{min}$ and upper $\lambda_{max}$ bounds of this distribution are given by
\begin{equation}
\lambda^{max}_{min} = \sigma^2 (1 + 1/Q \pm 2 \sqrt{1/Q}).
\label{lambda_bounds}
\end{equation}
In this expression $Q=T/N \ge 1$ and $\sigma^2$ is equal to the variance of the time series. The degree of departure of the largest eigenvalue $\lambda_1$ above $\lambda_{max}$ is a measure of the strength of correlations among the time series participating~\cite{kwapien2012,drozdz2001}.

Changes of the magnitude of the largest eigenvalue $\lambda_1$ in the rolling time-window of length $T=100$ trading days for the present $N=9$ versus the noise regime as set by $\lambda_{max}$ and $\lambda_{min}$ for these particular values of $T$ and $N$ are shown in Fig.~10. Furthermore, in the same Figure changes of the largest eigenvalue $\gamma_1$ of an analogous matrix composed of the $\rho_q(s)$ coefficients as defined by Eq.~\ref{rho.q} taking $q=2$ for $s=100$ are also shown.  Clearly, in both these measures the largest eigenvalues assume the lowest values in the period of interest, just between October 1990 and April 1994. At one point the $\lambda_1$ value even touches the border of purely random series. Thus, the scenario of the least correlated 9 companies here studied in this time period applies, indeed, which explains a narrow $f(\alpha)$ of the global 9-companies index.

\begin{figure}
\centering
\includegraphics[scale=0.4,keepaspectratio=true]{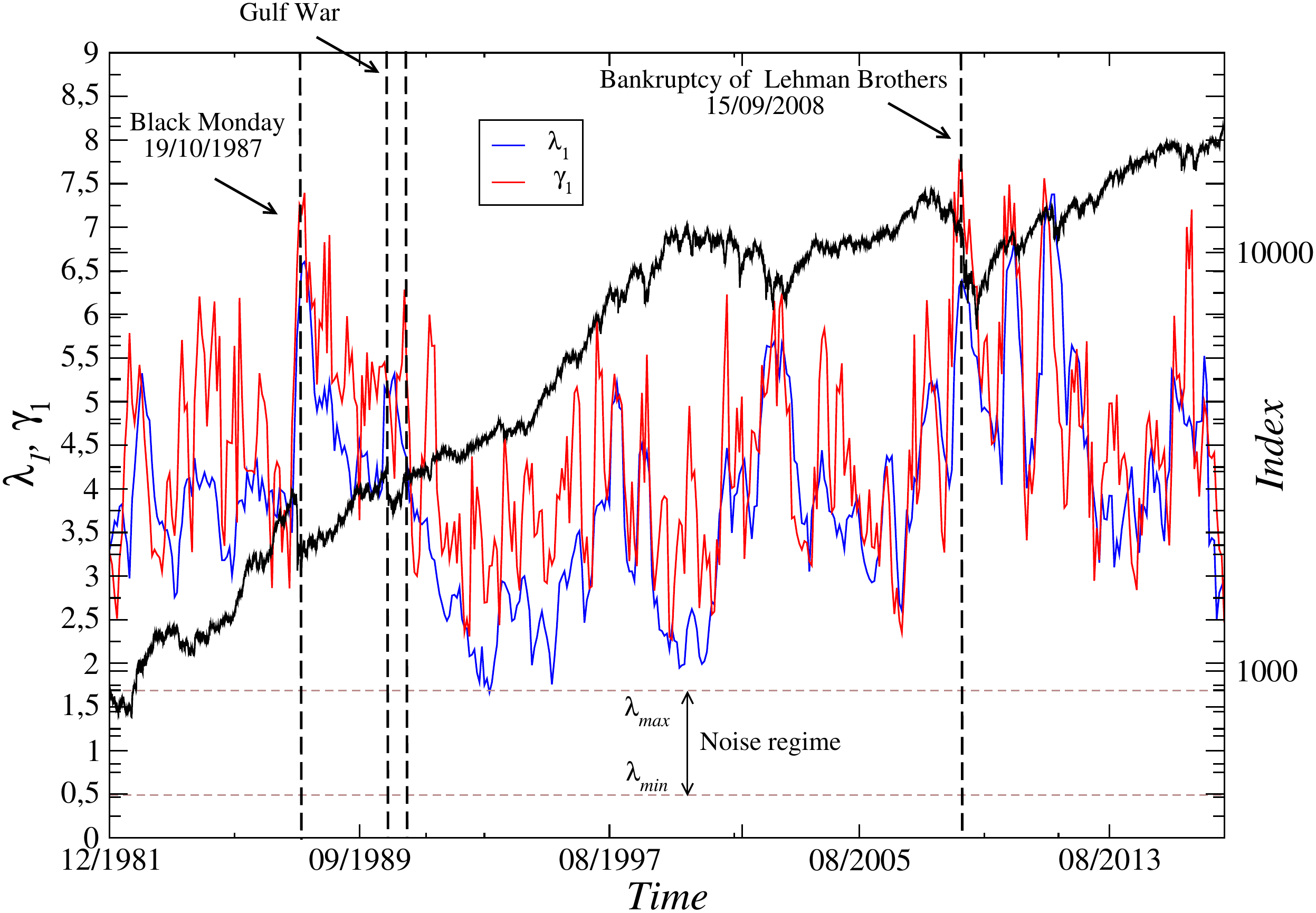}
\caption{Blue line displays the time-dependence of the largest eigenvalue $\lambda_1$ of the correlation matrix constructed from the time-series representing daily returns of the 9 companies of Fig.~9 in a rolling window of size $s=100$ trading days. Red line displays the largest eigenvalues $\gamma_1$ of an analogous rolling window matrix composed of the $\rho_q(s)$ coefficients as defined by Eq.~\ref{rho.q} taking $q=2$.}
\label{Fig10}
\end{figure}

\section{Conclusions}
\label{conclusions}
Quantification of the complex time series in terms of multifractality nowadays finds a multitude of applications in diverse areas. Thus far however majority of the related studies presented in the scientific literature limit themselves to a sole estimation of the singularity spectrum and, if found multifractal, it usually is treated as evidence of the hierarchical organization of such series and the width of such a spectrum is considered a measure of the degree of complexity involved. While this indicates some kind of a cascade-like, hierarchical organization indeed, in realistic cases such an organization is rarely uniform. The time series generated by natural processes may include many convoluted components with different hierarchy generators each, which results in asymmetry of the singularity spectra. Even more, contribution of such components may vary in time and this thus may introduce further dynamical variability. Definitely, the financial markets constantly functioning in evolving external conditions represent a natural candidate to become a subject of such effects. This can be anticipated to apply almost straightforwardly to the stock market indices as they by construction constitute an average (typically weighted but not always) of the prices of selected stocks representing different economy sectors thus not necessarily obeying the same multiscaling characteristics. The degree of correlations among such stocks is also known to depend on the global market phases. In the present paper, based on over half a century daily recordings of S\&P500 and NASDAQ, the two world leading stock market indices, it is shown that they reveal the multiscaling features which expressed in terms of the multifractal spectrum evolve through a variety of shapes whose changes typically appear correlated with the historically most significant events experienced by the world economy. From a more general perspective these results indicate that the form of the multifractal spectrum, and especially its departures from the model mathematical cases of the uniform cascades, contains richness of information that, if properly interpreted and potentially disentangled, may provide very valuable insight into the underlying dynamics which may be of crucial value for a more accurate modelling of the financial markets. Taking into consideration the effects exposed here may also be very helpful for market regulators and policy-makers in stabilizing markets as well as for a flexible portfolio optimization. 

Finally, the methodology introduced in subsection~\ref{IvsC} of relating the global (here index) multifractal spectrum to the corresponding multifractal spectra of subsystems (here companies) provides an appropriate quantitative tool with potential applications extending far beyond the financial context when various questions related to the so-called \textit{complexity matching}~\cite{west2008} are addressed and studied empirically as for instance those in a psychological/cognitive domain~\cite{stephen2008,stephen2011,stephen2012,abney2015,delignieres2016}. Differences between widths - as an example in Fig.~\ref{Fig9} shows - of such spectra reflect strength of the underlying \textit{complexity matching} between subsystems and this strength may vary in time. The weakest matching, for instance, corresponds to the period between October 1990 and April 1994. Furthermore, appreciating the relative changes in asymmetry of $f(\alpha)$ may allow to selectively scan the varying strength of such a matching for different ranges of fluctuations. Of course, as far as the world financial markets are concerned one may rely on observations only since, by their very nature, there exists no realistic possibility to set up the world financial experiments. Since phenomena belonging to the domain of social psychology definitely constitute a significant factor driving the markets a properly coordinated joint multidisciplinary effort may crucially help in understanding the cross-scale dependences and information flows in the financial markets and in other complex systems as well.  

\section*{Acknowledgements}
This research was supported in part by PLGrid Infrastructure.

\end{document}